\documentclass[a4paper, 12pt]{JHEP3}
\usepackage{latexsym,amsfonts,amsmath}
\usepackage{float}
\usepackage{amssymb}
\usepackage{graphicx}
\usepackage{epsfig}

\title{Fine Tuning in Quintessence Models with Exponential Potentials}

\author{Urbano Fran\c{c}a and Rogerio Rosenfeld\\ 
Instituto de F\'{\i}sica Te\'orica - UNESP\\ 
Rua Pamplona, 145\\ 
01405-900 S\~{a}o Paulo, SP, Brazil\\ 
E-mail: \email{urbano@ift.unesp.br}\\
E-mail: \email{rosenfel@ift.unesp.br}}

\abstract{We explore regions of parameter space in a simple exponential model
of the form $V =  V_0 \ e^{- \lambda \frac{Q}{M_p}}$ that are allowed by 
observational constraints. 
We find that the level of fine tuning in these models is not different from more
sophisticated models of dark energy. 
We study a transient regime where the parameter $\lambda$ has to be less than
$\sqrt{3}$  and the fixed point $\Omega_Q = 1$ has not been reached.
All values of the parameter
$\lambda$ that lead to this transient regime are permitted. We also 
point out that this model can 
accelerate the universe today even for $\lambda > \sqrt{2}$,  
leading to a halt of the present acceleration of the universe in the future 
thus avoiding the horizon problem.
We conclude that this model can not be discarded by current observations.
}

\keywords{Cosmology of Theories beyond the SM, Physics of the Early Universe}

\preprint{astro-ph/0206194 \\
IFT-P.039/2002 \\
PACS number: 98.80.Cq}

\begin{document}

\hyphenation{quin-tes-sen-ce}

\section{Introduction} \label{sec:introd}

The universe today is in a stage of accelerated expansion and we
still do not know exactly what is the physical mechanism that drives
this acceleration. Several alternatives have been
proposed, such as the old cosmological constant or the effect of a
field whose potential energy dominates nowadays the energy of the universe 
\cite{review}. This last proposal is usually called quintessence \cite{quintessence}.

One of the simplest models for quintessence is a scalar field rolling
down an exponential potential \cite{expo1}. Exponential potentials are natural
in theories with extra compact dimensions, such as Kaluza-Klein supergravity 
and superstring models \cite{models}.

Recently, it was pointed out the possibility of having exponential
quintessence with a temporary acceleration phase, avoiding
the horizon problem that appears in the context of string theory associated
with eternal acceleration \cite{endinflation}.

Exponential quintessence has been studied extensively in the
literature \cite{expo2,expo3} and usually discarded from limits on
big-bang nucleosynthesis. This is the motivation for introducing more complex
models with, {\it e.g.} two scalar fields \cite{orfeu} and double exponential
potentials \cite{double}. However, the arguments against the simple exponential
model are based on the assumption that 
a fixed point solution for the quintessence
evolution is reached early on in the universe. Therefore, one has to analyse
the case where the field have
not yet reached this fixed point. This is the goal of this article. We find regions of
parameter space for exponential quintessence that are allowed by
observational constraints and examine the level of fine
tuning required for realistic solutions.

\vskip 1.0cm

\section{Fixed point or fine tuning?}

We adopt the potential
\begin{equation} \label{eq:potexp}
V(Q) \: = \: V_0 \ e^{- \lambda \frac{Q}{M_p}} \ ,
\end{equation}
where $Q$ is the scalar quintessence field, $M_p = (8 \pi G)^{-1/2} $ is the 
reduced Planck mass
and $\lambda$ and $V_0$ are parameters of the potential.

The equation of motion for the quintessence field is given by:
\begin{equation}\label{eq:movimento}
\ddot{Q} + 3 H \dot{Q} \: = \: -\frac{d V}{d Q} \ ,
\end{equation}
where dot denotes differentiation with respect to time. Friedmann's 
equation for a flat universe is:
\begin{equation} \label{eq:friedmann}
H^2 \: = \: \frac{\dot{R}^2}{R^2} \: = \: \frac{1}{3 M_p^2}
\ \left( \rho_M + \rho_R + \frac{1}{2}\dot{Q}^2 + V(Q) \right) \ , 
\end{equation}
where $R$ is the scale factor and  $H$ is the Hubble parameter. 
For scalar fields, the energy density and 
pressure are respectively $\rho_Q = \frac{1}{2} \dot{Q}^2 + V(Q)$ and 
$p_Q = \frac{1}{2} \dot{Q}^2 - V(Q)$. The radiation and matter densities $\rho_R$ and 
$\rho_M$ evolve as:
\begin{equation}
\dot{\rho}_{M (R)} + n H \rho_{M (R)} \: = \: 0 \ ,
\end{equation}
with $n=3 (4)$ for non-relativistic (relativistic) matter. 

The acceleration $\ddot{R}$ of the universe is given by
\begin{equation} \label{eq:acceleration}
\frac{\ddot{R}}{R} \: = \: -\frac{1}{6 M_p^2} \sum_i  \rho_i \left( 1 + 3 \omega_i\right) ,
\end{equation}
where $i = M, R$ and $Q$, $\omega_M = 0$, $\omega_R = 1/3$ and
 $\omega_Q = p_Q / \rho_Q$ are the equation 
of state for matter, radiation and quintessence respectively.
The universe is accelerating today if $\omega_{Q0} < - \frac{1}{3}$, assuming 
that quintessence is currently the dominant component of the energy density. 
The index $0$ denotes the present epoch.

The dynamics of the quintessence field is obtained by solving
the non-linear, coupled differential equations (\ref{eq:movimento}) and 
(\ref{eq:friedmann}) with appropriate initial conditions.
This system of equations has stable fixed points ($fp$) for
$\lambda < \sqrt{n}$ with $\Omega_Q^{fp} = 1$ and $\lambda > \sqrt{n}$ 
with $\Omega_Q^{fp} = \frac{n}{\lambda^2}$ {\cite{expo3}}, where $n$ depends 
on which background component is more important when the field reaches the fixed point: 
$n=3 (4)$ for matter (radiation). 
The case $\lambda > \sqrt{n}$ is the selftuning (or scaling) scalar field discussed 
by Ferreira and Joyce {\cite{expo2}} and by Copeland, Liddle 
and Wands {\cite{expo3}}. In this case, however, either the 
system reaches 
its non-trivial fixed point early on in the universe and the value of 
$\Omega_{Q0}$ is too low or the quintessence field
energy density contributes too much to the total energy density in the early 
universe and spoils
big bang nucleosynthesis (BBN) predictions. Furthermore, in this selftuning 
case the universe 
is not accelerating today, since when the
fixed point is reached, $\omega_Q$ imitates the dominant background component equation of 
state (matter today) and consequently $\omega_{Q0} = \omega_M = 0$ 
{\cite{expo2,expo3}}. For these reasons, this solution can be discarded.

The case of interest is given by  $\lambda < \sqrt{3}$, in which the quintessence 
has not yet reached the fixed point regime today but will do it in the future.
In other words, $\Omega_Q$ is evolving from a small value 
(in order that quintessence does not spoil 
BBN) to its value at the fixed point, $\Omega_Q^{fp} = 1$. In this case, this is 
not a selftuning solution, but a tracking {\cite{steinhardt}} one, where $\omega_Q$  
is also frozen with the different value {\cite{expo3}} 
\begin{equation} \label{eq:omegafp}
\omega_Q^{fp} \: = \: \frac{\lambda^2 - 3}{3} \ ,
\end{equation}
being able to accelerate the universe today. Other regimes cannot 
explain BBN and/or the present acceleration of the universe, when one is using 
simple exponential potential.

In all quintessence models there arises the so-called 
cosmic coincidence problem: why is 
quintessence starting to dominate just now or, in other words, why is the 
value of quintessence 
energy density nearly equal to the matter energy density today? 
No quintessence model has solved this 
problem until now, since, in order to explain this coincidence, in general one 
adjusts the potential energy density to make the selftuning (or the tracking) 
field to freeze in a energy density 
that is of the order of the critical density today. 
In {\cite{steinhardt}}, for instance, two examples are
considered: $ V(Q) = K ^{( 4+ \alpha)} Q^{-\alpha}$ and 
$V(Q) = K^{4} \left[ \exp(M_p/Q) - 1 \right]$, where $M_p$ is the Planck 
mass and $K$ is a constant. For any given $V(Q)$, there is a family 
of tracking solutions parameterized by $K$, whose value is
fixed by the measured value of $\Omega_{M0}$. This means that the overall
scale of the potential for all quintessence models has to be of the order
of the present energy density. For this reason, $K$
is a free parameter used to fit current observations. Whether the order of
magnitude of the initial conditions and parameters can be obtained in 
reasonable physical
models is a very important question that has not been fully answered in a
satisfactory manner.

Hence, because the cosmic coincidence problem has not been explained by any quintessence
model, what one can do is to discuss the naturalness of the choice of initial
conditions and parameters.  Our goal in this article is exactly this: 
we study what is the size of the region in parameter space and 
how precisely adjusted these initial conditions and parameters must be in order to 
explain some observational constraints, that is,
what amount of fine tuning is necessary in simple exponential when  
one requires that it has not yet reached its fixed point regime.

\section{Relevant equations}

We will follow closely the approach described in Cline
{\cite{endinflation}}. Instead of integrating in time, we will use
another variable $u$ defined by:
\begin{equation} \label{eq:u}
u \: \equiv \: \ln (1 + z) \: = \: - \ln R/R_0 \ ,
\end{equation}
where $z$ is the redshift, and we take $R_0=1$. 
We also define new dimensionless variables:
\begin{equation} \label{eq:rescaling1}
\hat{Q} \: \equiv \: \frac{1}{\sqrt{3} M_p} \ Q \ , \qquad \qquad 
\hat{\rho}_i \: \equiv \: \frac{1}{3 M_p ^2 (H_0^m)^{2}}  \rho_i \ ,
\end{equation}
\begin{equation} \label{eq:rescaling2}
\hat{V} \: \equiv \: \frac{1}{3 M_p ^2 (H_0^m)^2}  \ V \ , \qquad  \qquad 
\qquad \hat{H} \: \equiv \: \frac{H}{H_0^m} \ ,
\end{equation}
where $H_0^m = 100 h_0 $ km s$^{-1}$ Mpc$^{-1}$ is the currently 
measured Hubble parameter and therefore:  
\begin{equation} 
3 M_p^2 (H_0^m)^2 \: = \: \rho_{c0}^m = \: 1.88 \times 10^{-29} \ h_0^2 \ 
\  \mathrm{g/cm^3}  \ .
\end{equation}
The dimensionless variables $\hat{\rho}_i$ and 
$\hat{V}$ are given in terms of the critical density. In our numerical
investigations, we adopt the following
values, for $h_0 = 0.7$:
\begin{equation} \label{eq:measured}
\hat{\rho}_{M0} \: =\:  \frac{\rho_{M0}}{\rho_{c0}^m} \: = \: 0.3 
\ \ \ \ \ \mathrm{and} \ \ \ \ \ 
\hat{\rho}_{R0} \: = \:  \frac{\rho_{R0}}{ \rho_{c0}^m} \: = \: 
4.3 \times 10^{-5} h_0^{-2} \: = \:  8.5 \times 10^{-5} \ .
\end{equation}
The variable $\hat{H}$ is determined by the numerical solution $H$ of 
Friedmann's equation 
(\ref{eq:friedmann}) divided by $H_0^m$. 

In terms of the new variables, the potential is given by 
\begin{equation} \label{eq:newpotential}
\hat{V}(\hat{Q}) \: = \: \hat{V}_0 \ e^{- \sqrt{3} \lambda \hat{Q}} \ ,
\end{equation}
and the differential equations read:
\begin{equation} \label{eq:movimentofinal}
\hat{H}^2 \hat{Q}''  -  \left( \hat{\rho}_R + \frac{3}{2}
\hat{\rho}_M  + 3 \hat{V} \right) \hat{Q}' + \frac{d
\hat{V}}{d \hat{Q}} \: = \: 0 \ ;
\end{equation}
\begin{equation} \label{eq:friedmannfinal}
 \hat{H}^2 \: = \: \frac{\hat{ \rho}_M + \hat{\rho}_R+
 \hat{V}}{ 1 - \frac{1}{2}\hat{Q'}^2  } \ ;
\end{equation}
where prime denotes differentiation with respect to $u$.
This corrects a typographical error in Cline {\cite{endinflation}},
where he writes $\hat{V}$ instead of $ 3 \hat{V}$ in equation
(\ref{eq:movimentofinal}). Notice that because the universe 
is flat, one is able to isolate $\hat{H}$ in Friedmann's equation. 

These are supplemented by the solutions for the evolution of matter
and radiation
\begin{equation} \label{eq:densidademateria2}
\hat{\rho}_M \: = \: \hat{\rho}_{M0} \ e ^ {3u} \ ,
\end{equation}
and
\begin{equation} \label{eq:densidaderadiacao2}
\hat{\rho}_R \: = \: \hat{\rho}_{R0} \ e ^ {4u} \ ,
\end{equation}
where $\rho_{R0}$ and $\rho_{M0}$ are the radiation and matter energy densities today.
The quintessence energy density is given by:
\begin{equation} \label{eq:densidadequinte2}
\hat{\rho}_Q \: = \: \frac{1}{2} \ \hat{H}^2 \hat{Q'}^2 + \hat{V} \ .
\end{equation}
With these new variables, one can also write
\begin{equation} \label{eq:parametrodens}
\Omega_i \: = \: \frac{\rho_i}{\rho_c} \: = \:
\frac{\rho_i}{3 M_p^2 H^2} \: = \: \frac{\hat{\rho}_i}{\hat{H}^2} \ .
\end{equation}
that always satisfy
\begin{equation} \label{eq:um}
\Omega_R + \Omega_M + \Omega_Q \: = \: 1 \ .
\end{equation}

From the equation above, one can see that the values 
of $\Omega_{M0}$, $\Omega_{R0}$ and $\Omega_{Q0}$ 
(since the universe is flat) depends on the solution of 
Friedmann's equation. The values  (\ref{eq:measured}) used to solve 
the equations will correspond to the measured values of 
$\Omega_{M0}$ and $\Omega_{R0}$ only if $H_0 = H_0^m $.

The equation of state for quintessence is given by:
\begin{equation} \label{eq:eqestadofinal}
\omega_Q \: = \: \frac{\dot{Q}^2 - 2 V}{\dot{Q}^2 + 2 V} \: =
 \: \frac{ \hat{H}^2
\hat{Q'}^2 - 2 \hat{V}}{\hat{H}^2 \hat{Q'}^2 + 2 \hat{V}} \ .
\end{equation}

\section{Numerical Results}

\subsection{Parameters} \label{ssec:parameters}

As stated before, in all quintessence 
models there is an overall constant ($\hat{V_0}$ in our case) that is 
determined by the fact that the major contribution to
the energy of the field today must come from the potential term, and that the energy
density of quintessence is approximately equal the present measured critical density 
(more precisely, the critical energy density minus the matter plus radiation energy densities). 
This work does not aim to discuss the naturalness of such a choice of 
$\hat{V_0}$. 
We simply follow the same approach used in all quintessence models to ``solve'' 
the cosmic coincidence problem.

From equation ({\ref{eq:eqestadofinal}}) one sees that the pressure-to-density 
ratio (equation of state) 
has a current negative value only if the major contribution for the energy
density of the quintessence field comes from the potential
term. The observational fact that $\Omega_{Q0} \approx \rho_{c0}^m$ implies that
\begin{equation}
V(0) \: = \: V_0 \ e^{- \sqrt{3} \lambda \hat{Q}_0} \: 
\approx \: \rho_{c0}^m \: \approx \:  (10^{-3} \mbox{eV})^4 \ ,
\label{eq:V0}
\end{equation}
where $V(0)$ denotes the present value of the potential energy density. The value of 
$V_0$ depends on the value of $\hat{Q}_0$.
As long as the initial condition $\hat{Q'}_i$ is small (and we will see that this is the case from
the equipartition of energy) and the field 
has not yet reached the fixed point, one can see that 
$\hat{Q}_0 \approx \hat{Q}_i$, as figure \ref{fig:qu} illustrates. 
Therefore, the values of $V_0$ and $\hat{Q}_i$ are
related by equation ({\ref{eq:V0}}). 

In this sense, the only free parameter in this model is $\lambda$. As mentioned above, 
we will be interested in the range
$0 \leq  \lambda  <  \sqrt{3}$, since this region is the only one able to 
explain all observational constraints.

\vspace{1.0cm}

\FIGURE{\epsfig{figure =q_u.eps, scale=0.45}
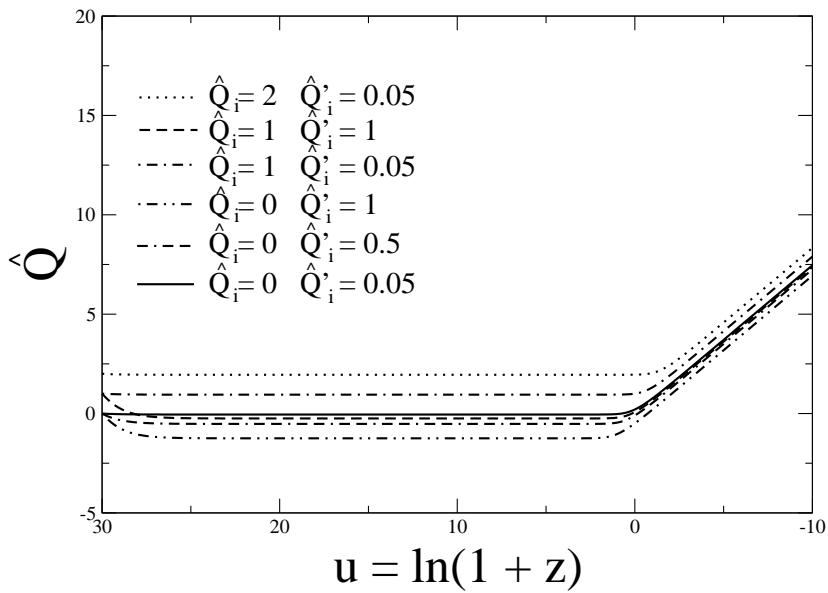
\caption{Value of the field $\hat{Q}$ as function of 
$u$. For the initial conditions showed, 
the field has approximately the same features. The field remains almost 
constant until it reaches the fixed point regime, when it starts
to roll-down the potential with a constant derivative in $u$. The initial conditions 
used along the text are showed by solid line. These curves were made with $\hat{V}_0 = 1$ 
and $\lambda = 1.3$.}\label{fig:qu}}

\subsection{Initial Conditions}

In order to solve the differential equations, two extra parameters
are needed, namely the initial conditions $\hat{Q}_i$ and $\hat{Q'}_i$. 
We are interested in the case in which the field has not yet entered
in the fixed point regime. In this case, the field has today almost
the same value it had initially, and one is able to relate $V(0)$ and 
$\hat{Q}_i$, in such a
way that  $\hat{Q}_i$ can be 
absorbed in the definition of $\hat{V}_0$. For this reason, we will take, 
with no loss of generality, $\hat{Q}_i = 0$. We will discuss later 
the role of $\hat{Q}_i$ in the solution. 
In fact, changing the value of $\hat{Q}_i$ 
just corresponds to a rescaling of the problem.

The freedom in the choice of 
$\hat{Q}_i$ also comes from the fact that  the potential 
energy does not contribute to the initial value 
of density parameter of the field, $\Omega_{Q,i}$. This can be seen by noticing
that $V(0)$ is of the order of the critical density today (see equation 
({\ref{eq:V0}})) and that our initial conditions 
are taken at $ z \approx 10^{13}$. This implies that $\hat{V_0}$ is at least 
39 orders smaller than the energy density scale (critical density) 
of that epoch, since $\rho_c \propto 
H^{2} \propto R^{-3} \ (R^{-4})$ 
during matter (radiation) domination epoch. The 
initial energy of the field is then in the form of kinetic energy, or
\begin{equation} \label{eq:kineticenergy}
\Omega_{Q, i} \: \approx \: \frac{1}{2} \hat{Q'}_i ^ 2 \ .
\end{equation}
We have assumed  
a flat universe, which implies that $| \hat{Q'} | \leq \sqrt{2}$. The equation 
above shows what would be needed in order to have the limit case $| \hat{Q'} | = \sqrt{2}$ : 
the dominant energy of the universe must be the kinetic energy of the field.   

Natural initial conditions from equipartition of energy after inflation 
suggests that $\Omega_{Q, i}  \simeq 10^{-3}$ 
{\cite{steinhardt}}. From equation ({\ref{eq:kineticenergy}}) one has then 
$\hat{Q'}_i \approx 0.05$. 

Therefore, the most likely set of initial conditions are 
the values $\hat{Q}_i = 0$ and $\hat{Q'}_i = 0.05$. Nevertheless, 
we will also study the effect of  taking all the possible set of 
initial conditions, 
namely $\hat{Q}_i \geq 0$ and $0 \leq  |\hat{Q'}_i| \leq \sqrt{2}$.  
 
\subsection{Constraints} \label{ssec:constraints}

We evaluate the equations and demand that the solutions must satisfy 
some observational constraints. 
The first is given by nucleosynthesis. 
Nucleosynthesis predictions claims {\cite{limit}} that at 
$95$\% confidence level:
\begin{equation} \label{eq:nucleosynthesis}
\Omega_Q (1 \mbox{MeV} \approx z=10^{10}) \: \leq \: 0.045 \ .
\end{equation}
Another observational constraint is given by the quintessence density parameter. 
Observations from cosmic background radiation
anisotropy indicate that the universe is flat. A set of complementary 
observations indicates that 
$\Omega_{M0} = 0.3 \pm 0.1$ {\cite{observations}}. Thus,  
\begin{equation} \label{eq:density}
\Omega_{Q0} \: = \:  0.7 \pm 0.1 \ .
\end{equation}
The uncertainty on $\Omega_{M0}$ implies that
there is an uncertainty on the determination of
$\hat{V}_0$. This is the reason why one can study a region on parameter space 
$(\hat{V}_0, \lambda)$ in spite of the fact that $\hat{V}_0$ is not a free parameter.   
    
The last observational constraint considered here is the present 
quintessence equation of state  \cite{observations}:   
\begin{equation} \label{eq:state}
-1 \: \leq \: \omega_{Q0} \: \leq \:  -0.6 \ .
\end{equation}
The fact that the universe is accelerating today, as will be shown in
figure {\ref{fig:wu}}, does not imply that it will accelerate forever. The 
present value of equation of state of quintessence in the cases studied in this work is 
temporary, since the equation of state is frozen only  when the field reaches
the fixed point regime, what will happen in near future.
  
\subsection{Results and Discussion}

We solve numerically the coupled differential equations 
({\ref{eq:movimentofinal}},{\ref{eq:friedmannfinal}}) using equations 
(\ref{eq:newpotential},{\ref{eq:densidademateria2}},{\ref{eq:densidaderadiacao2}}).
The effect of all possible initial conditions
will be discussed later. First, however, the particular choice (the most likely one) of initial 
conditions $\hat{Q}_i = 0$ and 
$\hat{Q'}_i =  0.05$ will be studied.

For this set of initial conditions, the region of parameter 
space\footnote{It is useful to recall that 
we are interested in models with $\lambda  <  \sqrt{3}$ (
see section {\ref{sec:introd}}).} able to satisfy all observational 
constraints, contrary to common belief, is reasonable, varying from 
$\lambda  =  0$ to $\lambda  \approx 1.7$
and from $\hat{V}_0 \approx 0.5$ to $\hat{V}_0 \approx 3.0$. In other words,
{\bf all} possible values of $\lambda$ in the tracking regime are able to 
satisfy the present 
observational constraints. The results are shown in figure {\ref{fig:lambdav0}}.


\FIGURE{\epsfig{figure=Lambda_V_0_005.eps, scale=0.45}
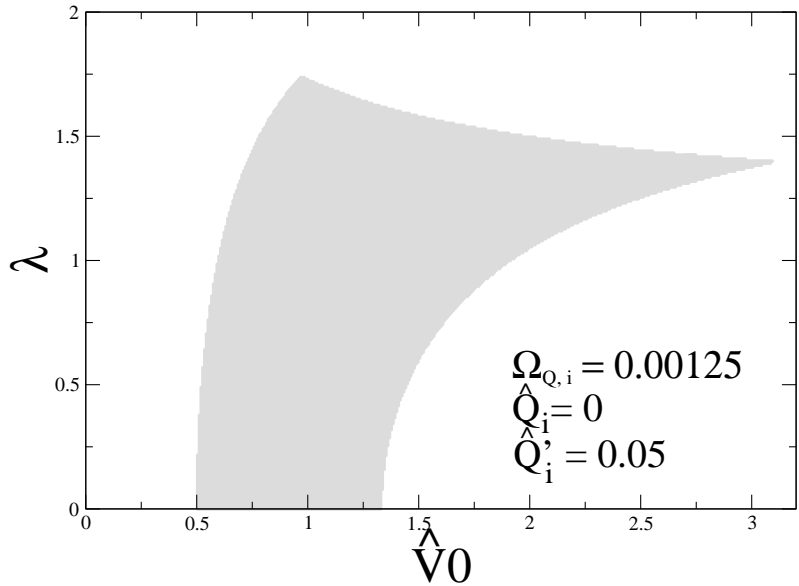
\caption{Region of parameter space that satisfies 
all observational constraints discussed in 
subsection {\ref{ssec:constraints}}. There is a 
reasonable region of parameters of the exponential potential that can 
explain all observations. In fact, all values of $\lambda$ that produce the
tracking solutions satisfy the constraints. The uncertainty on $\hat{V}_0$ comes from 
the uncertainty on $\Omega_{M0}$.}\label{fig:lambdav0}}

Figures {\ref{fig:rhou}}, {\ref{fig:wu}} and {\ref{fig:Omegau}} shows respectively
how the energy densities, the equation of state of quintessence and density parameters 
varies with $u$. Initially,
quintessence contributes to a small fraction of energy of the universe and 
decreases as $R^{-6}$, dominated by the kinetic term ($\omega_Q = 1$), 
faster than matter and radiation densities. When the potential term becomes 
important, there is a rapid change in the equation of state from $\omega_Q = 1$ 
to  $\omega_Q = -1$ and the 
quintessence density freezes until today, when it becomes dominant. Afterwards, 
the quintessence reaches the fixed point regime, 
which is characterised by the quintessence density parameter going from as small value 
at  $u \approx 2.5$ and reaching the fixed point, $\Omega_Q^{fp} = 1$, when 
$u \approx -2.5$. In the fixed point regime (tracking solution), the ratio 
between kinetic and potential energy
densities becomes constant
and consequently $\omega_Q^{fp}$ is given by the equation (\ref{eq:omegafp}). 
In this regime, the energy density  
decreases as $R^{- 3(1+\omega_Q^{fp})} = R^{- \lambda^2}$.   

Figure \ref{fig:wu} shows the behavior of equation of state for
various parameters. Initially they have the same 
behavior, but they evolve differently after 
quintessence enters the fixed point regime, because different 
parameters correspond to different frozen ratios between kinetic and potential 
energy densities. $\lambda$ is the parameter that determines what 
value $\omega_Q^{fp}$ will be frozen in: small values of $\lambda$ corresponds to 
small values of $\omega_Q^{fp}$ and vice versa. In particular, for
$\lambda > \sqrt{2} $ the universe stops to 
accelerate in the tracking solution, since $\omega_Q^{fp} > -\frac{1}{3}$ in these cases. 
For small values ($\lambda \lesssim 0.15$) the quintessence behaves 
like a cosmological constant in the tracking solution.


\FIGURE{\epsfig{figure=rhoq_u_v1_l1.eps, scale=0.45}
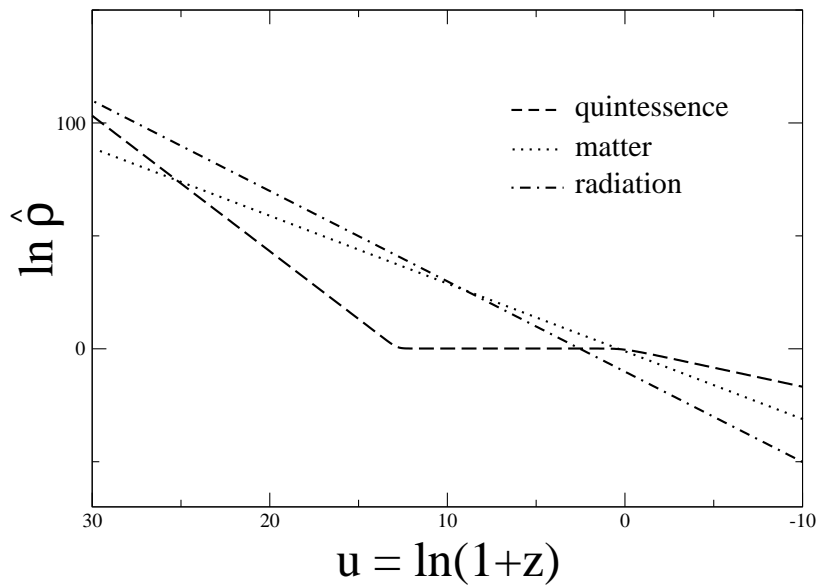
\caption{Energy density as function of $u$ for quintessence, 
matter and radiation for a typical 
solution. Initially quintessence density 
decreases faster than matter and radiation. When the potential
term becomes important, 
$\hat{\rho}_Q$ freezes.  When the fixed point regime (tracking solution) is 
reached, the field changes again its behavior and decreases as $R^{- \lambda^2}$, 
lower than matter. Plot made with parameters $\hat{V}_0  =  1$ and 
$\lambda  =  1.3$.}\label{fig:rhou}}


\FIGURE{\epsfig{figure= w_u.eps, scale=0.45}
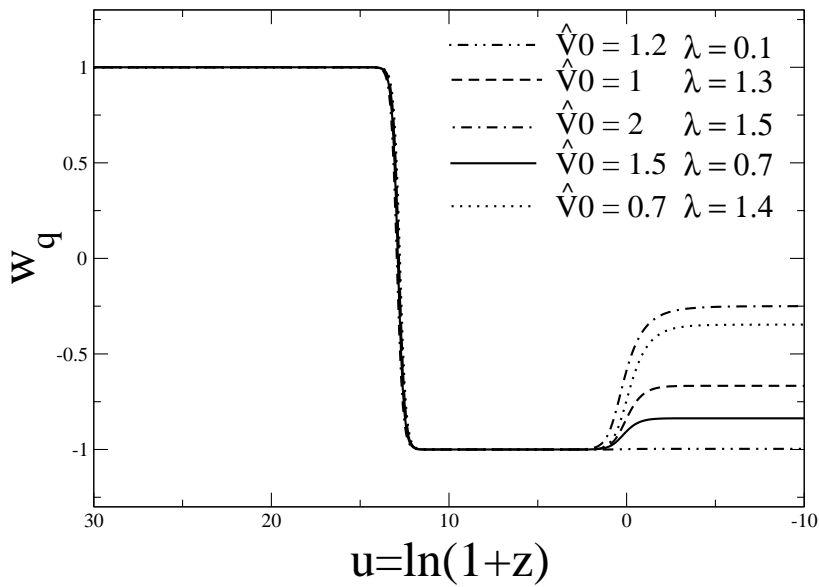
\caption{Equation of state as function of $u$ for 
various parameters. Initially all
have the same behavior. When quintessence reaches its fixed point, 
each solution is frozen in a specific value, depending mainly on $\lambda$: 
small values of $\lambda$ correspond to smaller values of $\omega_Q^{fp}$ and 
vice versa. For a small $\lambda$, quintessence behaves in the fixed point regime 
like a cosmological constant. For $\lambda > \sqrt{2}$, the universe
does not accelerate in this regime.}\label{fig:wu}}


\FIGURE{\epsfig{figure=Omega_u.eps, scale=0.45}
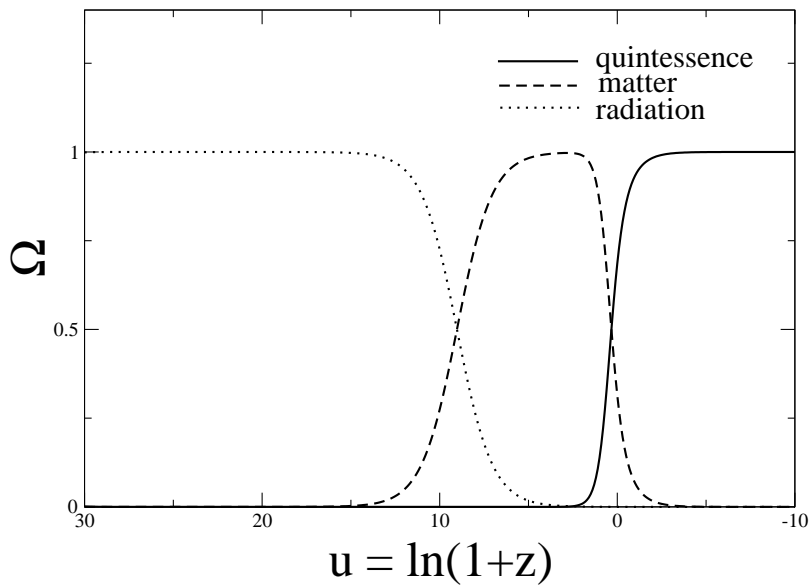
\caption{Density parameters of quintessence, matter and radiation. The fixed point 
regime is characterised by quintessence density parameter equal to unity ($\Omega_Q^{fp} = 1$). 
The fixed point solution was not yet reached today, which is a 
transition epoch. The parameters used were
 $\hat{V}_0 =  1$ and $\lambda =  1.3$.}\label{fig:Omegau}}

The correlation between 
the free parameter $\lambda$ and  the equation of state today
is shown in figure {\ref{fig:wlambda}}. Notice that there is 
a degeneracy for $\lambda \gtrsim 0.6$, namely, different 
values of $\lambda$ generate the same $\omega_{Q0}$. This degeneracy
is mainly due the fact that the present value of equation of state is not its 
value in the fixed point regime, since the tracking solution was not yet reached.
In the fixed point regime the degeneracy does not exist, since equation 
(\ref{eq:omegafp}) is satisfied. 


\FIGURE{\epsfig{figure=w_lambda.eps, scale=0.45}
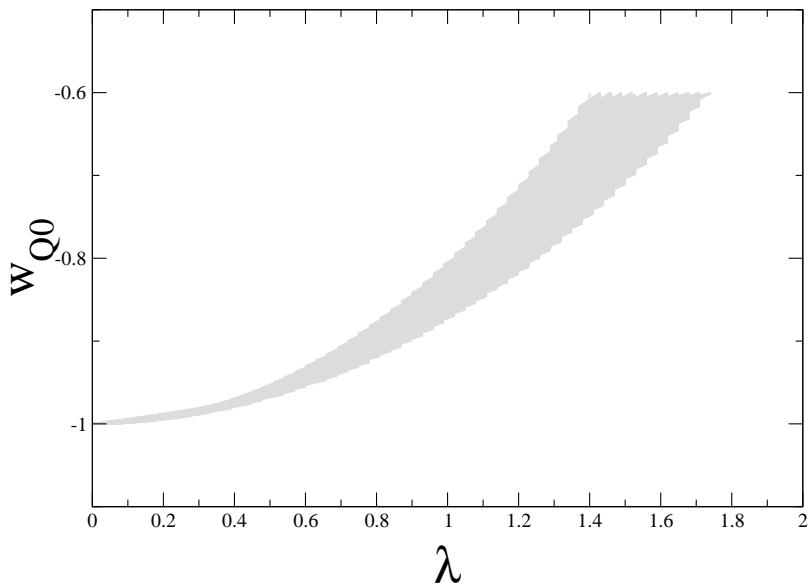
\caption{Allowed region of ($\lambda,\omega_{Q0}$)
space. Notice that 
$\omega_{Q0} \leq -0.6$, according to constraint (\ref{eq:state}). This plot
is independent on values of $\hat{V}_0$ and $\hat{Q}_i$ and changes only for ``high'' 
values of $\hat{Q'}_i$ ($\hat{Q'}_i \gtrsim 1.25$). The fact that different values of $\lambda$
are able to give the same $\omega_{Q0}$ comes from the fact that today the fixed point
regime was not yet reached.}\label{fig:wlambda}}

With better measurements
of $\omega_{Q0}$, one could put severe constraints on the 
exponential potential model, specially if $ \omega_{Q0}  \lesssim  - 0.85$. 
This is the region of low $\lambda$ and, as it was commented before, in this
region quintessence behaves as a cosmological constant today. 
It is important to realize that 
the figure \ref{fig:wlambda} remains the same for almost all sets of values 
of $\hat{V}_0$, $\hat{Q}_i$ and $\hat{Q'}_i$ that are able to 
produce a tracking solution we are 
interested here\footnote{$\hat{Q}_i \geq 0$, $0 \leq | \hat{Q'}_i | \leq \sqrt{2}$ and 
$\hat{V}_0$ determined in the way discussed in subsection {\ref{ssec:parameters}}.},
since the value of equation of state depends mainly on $\lambda$. 
In fact, in order to have this plot unchanged, it is enough
that exists  a considerable region of $\lambda$ in the parameter space ($\hat{V}_0, \lambda$).
Later will be shown that the existence of this region only depends on 
$\hat{Q'}_i$. Only for values of $\hat{Q'}_i \gtrsim 1.25$ this plot
will be changed. 

Another important aspect is the dependence of these results on $\hat{Q}_i$. 
Figure {\ref{fig:lambdav1}} shows the 
region of parameter space that satisfies all observational constraints when one 
uses as initial conditions 
$\hat{Q}_i  =  1$ and $\hat{Q'}_i  =  0.05$. When $\hat{Q}_i  \neq  0$, the region of 
parameter space is deformed in $\hat{V}_0$ by a 
factor of $e^{ \sqrt{3} \lambda \hat{Q}_i}$, because in these models 
$\hat{V}_0$ can always be rescaled as
$\hat{v}_0  =  \hat{V}_0 e^{- \sqrt{3} \lambda \hat{Q}_i }$,
since $\hat{Q}_0 \approx \hat{Q}_i$. 
Hence, in this sense, choosing another value of 
$\hat{Q}_i$ corresponds to just a rescaling of the ``old'' region.


\FIGURE{\epsfig{figure=Lambda_V_1_005.eps, scale=0.45}
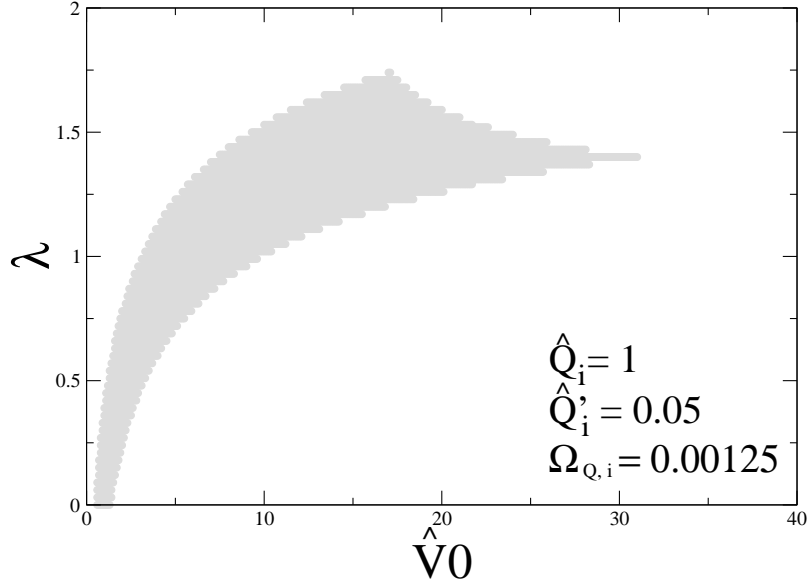
\caption{Region of parameter space that satisfies all observational 
constraints, when one is using 
as initial conditions $\hat{Q}_i  =  1$ and $\hat{Q'}_i  =  0.05$. 
Note that using $\hat{Q}_i  \neq  0$ 
just corresponds to rescale the parameter space in $\hat{V}_0$ by
a factor of $e^{ \sqrt{3} \lambda \hat{Q}_i}$.}\label{fig:lambdav1}}

 As it was seen in equation 
({\ref{eq:kineticenergy}}), the initial condition $\hat{Q'}_i$ corresponds to the 
initial value  of density parameter of quintessence. The most likely value
is $\hat{Q'}_i  =  0.05$ because of the energy equipartition after 
inflation. However, one can argue that choosing this value could also be
a fine tuning. For this reason, it is important to explore what happens with the allowed region of the
parameter space if $\hat{Q'}_i$ has a different value, keeping in mind that
$0 \leq |\hat{Q'}_i| \leq \sqrt{2}$. The results are shown 
in figure {\ref{fig:lambdav2}}.

The region of parameter space becomes smaller in $\hat{V}_0$
when $\hat{Q'}_i$ increases. This happens because the field has to satisfy
the constraint (\ref{eq:nucleosynthesis}). When the density parameter of quintessence is large 
initially,  in general the density parameter of quintessence
contributes too much to the total energy density during the BBN epoch, and a smaller
region of parameter space is able to satisfy the observational constraints. 
Nevertheless, a reasonable region of parameter space still exists for almost all values of
$\hat{Q'}_i$. In fact, the region of parameter space only vanishes 
when $\Omega_{Q,i}  \gtrsim  0.75$. In other words, for almost all 
possible values of $\hat{Q'}_i$ in a flat universe there still is a significant
region of the parameter space able to satisfy all observational constraints.


\FIGURE{\epsfig{figure=Lambda_V.eps, scale=0.45}
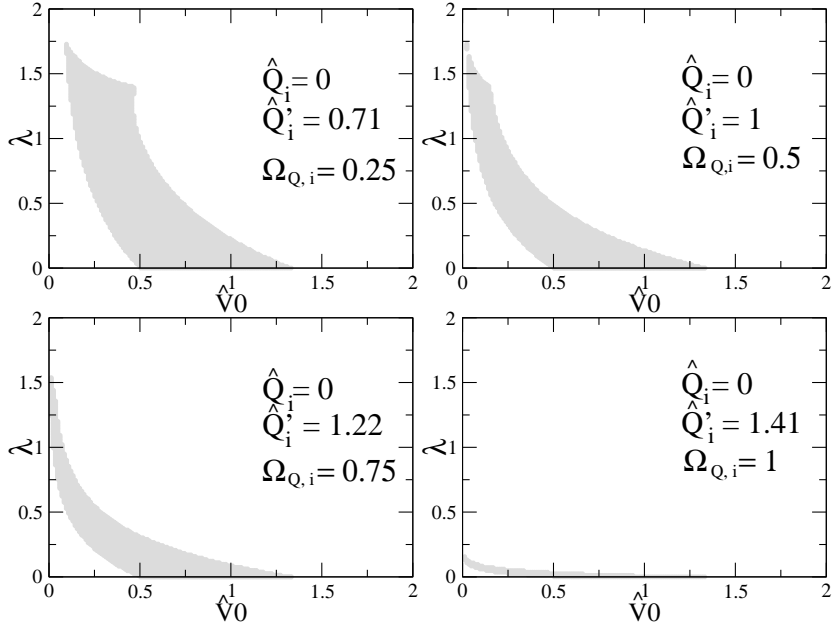
\caption{Regions of parameter space for various $\hat{Q'}_i$ different from that of 
inflationary equipartition. Since the universe is flat, 
$|\hat{Q'}_i| \leq \sqrt{2}$. For almost all possible values of $|\hat{Q'}_i|$
a significant region on parameter space still exists. It only vanishes for
$\Omega_{Q,i} \gtrsim 0.75$.
}\label{fig:lambdav2}}

\section{Conclusion}

In this work we have studied the simplest quintessence model with 
an exponential potential. This potential is usually discarded because
it cannot satisfy simultaneously all observational constraints. This 
is true in a regime where the 
field has already reached its fixed point regime. We have studied this model in
a regime where the field has not yet reached its fixed
point regime (tracking solution) today. We have shown that, contrary to common
belief, this potential is able to satisfy all observational constraints for a 
reasonable region of parameter space ($ \hat{V}_0, \lambda $). 

We have also shown that the resulting parameters and initial 
conditions are not unnatural. 
On the contrary, we showed that these parameters and initial conditions 
are not less natural than that used for other 
models of quintessence. For almost all possible values of 
$\hat{Q'}_i$ there still is a significant allowed region 
of parameter space. $\hat{V}_0$ is determined by the measured value 
of $\Omega_{M0}$, the same feature used in all quintessence models. 
$\hat{Q}_i$ does not affect qualitatively the region of parameter space, 
since in the regime in which the field
has not entered its tracking solution $\hat{Q}_i$
and $\hat{V}_0$ are related, and consequently $\hat{Q}_i$ just rescales the problem. 
The only free parameter of this model is $\lambda$, which determines the behavior of the field.
All values of $\lambda$ that produce the tracking solution satisfy the observational constraints.
Depending on $\lambda$, the universe may or may not accelerate forever.

The allowed regions that we found are essentially due to
present experimental uncertainties: the region of $\hat{V}_0$ is due the uncertainties
on measured  value of  $\Omega_{M0}$ and the region on $\lambda$ arises
from the observational uncertainties on $\omega_{Q0}$. If the uncertainties
were reduced, these regions would also be reduced. 
In this case, one would be able to determine the values of the parameters of 
the simple exponential potential using observations. 

This potential
cannot be discarded by any of the constraints discussed here, even if better 
measurements were made. The only way to discard this potential based on the constraints
discussed in this work would be if constraint of BBN (\ref{eq:nucleosynthesis}) had been more
stringent. For example, if one had showed that $\Omega_Q(1 \mathrm{MeV}) \leq 10^{-6}$, 
probably the region of parameter space would become smaller, in such way that could not be possible 
to explain the measured value of $\omega_{Q0}$, for example. Another way to discard this potential
is to include another constraint as, for instance, the value of $\Omega_Q$ during 
the epoch of formation of structure or in the last scattering surface, and verify
that this model is not able to satisfy all constraints simultaneously. 
Therefore, we showed that at the moment there is
no reason to discard the exponential potential or to consider it less natural 
than any other quintessence model.

\acknowledgments

This work was supported by Funda\c{c}\~{a}o de Amparo \`{a} Pesquisa do
Estado de S\~{a}o Paulo~(FAPESP), grant 01/11392-0 and Conselho Nacional de Desenvolvimento
Cient\'{\i}fico e Tecnol\'{o}gico~(CNPq). We would like to thank Orfeu Bertolami and Edmund J.
Copeland for their relevant comments.


\end{document}